\documentclass[aps,10pt,reprint,groupedaddress,superscriptaddress,nofootinbib]{revtex4-2}
\usepackage[usenames,dvipsnames]{xcolor}
\usepackage{amssymb}
\usepackage{amsmath}
\usepackage{dsfont}
\usepackage{txfonts}

\usepackage{graphicx}
\usepackage{mathptmx, textcomp}
\usepackage{braket,amsfonts}
\usepackage[dvipsnames]{xcolor}
\usepackage{upgreek}
\usepackage{txfonts}
\usepackage{braket}
\usepackage[english]{babel}
\usepackage{blindtext}
\usepackage{physics}
\usepackage{enumitem}
\usepackage[normalem]{ulem}
\usepackage[hidelinks, colorlinks=true,allcolors=blue]{hyperref}
\usepackage[all]{hypcap}
\usepackage{nicefrac}
\usepackage{float}
\usepackage{float}
\usepackage{tikz}
\usepackage{siunitx}

\setlist[itemize]{itemsep=0pt}

\newcommand{\omegaP}{\omega_{\perp}}

\begin{document}

\title{Experimental Observation of Single- and Multi-Site Matter-Wave Solitons\\ in an Optical Accordion Lattice}
 
\author{Robbie Cruickshank}
\affiliation{Department of Physics and SUPA, University of Strathclyde, Glasgow G4 0NG, United Kingdom }
\author{Francesco Lorenzi}
\affiliation{Dipartimento di Fisica e Astronomia ``Galileo Galilei'', Università di Padova, Via Marzolo 8, 35131 Padova, Italy}
\affiliation{Istituto Nazionale di Fisica Nucleare, Sezione di Padova, Via Marzolo 8, 35131 Padova, Italy}
\author{Arthur~La~Rooij}
\affiliation{Department of Physics and SUPA, University of Strathclyde, Glasgow G4 0NG, United Kingdom }
\author{Ethan F.~Kerr}
\affiliation{Department of Physics and SUPA, University of Strathclyde, Glasgow G4 0NG, United Kingdom }
\author{Timon Hilker}
\affiliation{Department of Physics and SUPA, University of Strathclyde, Glasgow G4 0NG, United Kingdom }
\author{Stefan Kuhr}
\affiliation{Department of Physics and SUPA, University of Strathclyde, Glasgow G4 0NG, United Kingdom }
\author{Luca Salasnich}
\affiliation{Dipartimento di Fisica e Astronomia ``Galileo Galilei'', Università di Padova, Via Marzolo 8, 35131 Padova, Italy}
\affiliation{Padua QTech Center, Universita di Padova, Via Marzolo 8, 35131 Padova, Italy}
\affiliation{Istituto Nazionale di Fisica Nucleare, Sezione di Padova, Via Marzolo 8, 35131 Padova, Italy}
\affiliation{Istituto Nazionale di Ottica del Consiglio Nazionale della Ricerche, Unit of 
Sesto Fiorentino, Via Carrara 1, 50019 Sesto Fiorentino, Italy}
\author{Elmar Haller}
\affiliation{Department of Physics and SUPA, University of Strathclyde, Glasgow G4 0NG, United Kingdom }

\date{\today}

\begin{abstract}  
We report the experimental observation of discrete bright matter-wave solitons with attractive interaction in an optical lattice. Using an accordion lattice with adjustable spacing, we prepare a Bose-Einstein condensate of cesium atoms across a defined number of lattice sites. By quenching the interaction strength and the trapping potential, we generate both single-site and multi-site solitons. Our results reveal the existence and characteristics of these solitons across a range of lattice depths and spacings. We identify stable regions of the solitons, based on interaction strength and lattice properties, and compare these findings with theoretical predictions. The experimental results qualitatively agree with a Gaussian variational model and match quantitatively with numerical simulations of the three-dimensional Gross–Pitaevskii equation, extended with a quintic term to account for the loss of atoms. Our results provide insights into the quench dynamics and collapse mechanisms, paving the way for further studies on transport and dynamical properties of matter-wave solitons in lattices.
\end{abstract}

\maketitle

%**********************************************
% Intro
%**********************************************

Bright solitons are localized wave packets that propagate without spreading over a low-intensity background in a nonlinear medium~\cite{kivshar1989dynamics}. They arise when the nonlinear self-focusing in the medium balances the dispersive spreading of the wave. Bright solitons have been observed in various physical systems, including optical fibers~\cite{mollenauer1980}, fluids~\cite{grimshaw2007}, and  quantum gases~\cite{abdullaev2005}. In particular, Bose–Einstein condensates with attractive interactions have been instrumental in studying matter-wave solitons in homogeneous systems, experimentally demonstrating the formation~\cite{bradley1997,strecker2002, khaykovich2002}, collapse~\cite{sackett1998, gerton2000, donley2001}, and collisions~\cite{nguyen2014} of bright solitons. 

Based on a seminal theoretical insight by Davydov~\cite{davydov1973}, solitons have also been studied in systems with periodic potentials. Such ``lattice solitons'' \cite{SolitonTerm} arise in wide range of systems \cite{malomed2006a}, including molecular chains~\cite{davydov1973, kruglov1984}, nonlinear optical waveguide arrays~\cite{eisenberg1998, morandotti1999, fleischer2003b, malomed2019}, and quantum gases in optical lattices~\cite{trombettoni2001a, ostrovskaya2003matter, eiermann2004}. They exist in both one-dimensional and two-dimensional geometries~\cite{ahufinger2004a, baizakov2003, morandotti1999, fleischer2003} and are predicted to exhibit intricate transport properties~\cite{trombettoni2001a, ahufinger2004a, brazhnyi2011, franzosi2011}. However, despite considerable theoretical interest~\cite{trombettoni2001a, abdullaev2001, efremidis2003, sakaguchi2005, salasnich2007,  maluckov2008, kartashov2011, rubbo2012, naldesi2019}, the experimental realization of lattice solitons with attractive matter waves has remained an open challenge.

Lattice solitons can be classified into single-site (SS) and multi-site (MS) solitons, which extend over different numbers of lattice sites, as well as on-site and off-site solitons, which are centered directly on sites or between them~\cite{salasnich2007}. In this work, we provide an experimental demonstration of both single-site and multi-site solitons of attractively interacting matter waves. These solitons form near the center of the Brillouin zone with energies below the lowest lattice band~\cite{efremidis2003, louis2003}. This is in contrast to gap solitons with repulsive interactions \cite{eiermann2004, mitchell2021} that appear in the energy gap near the band edge. We investigate the solitons' stability and decay dynamics, and compare our findings with theoretical predictions. A key element of our experimental approach is an accordion lattice with variable lattice spacing $d_\text{L}$~\cite{fallani2005, alassam2010, ville2017}, which serves three primary roles: the preparation of an initial wave packet in a given number of sites, the study of solitons for varying lattice spacing, and a magnification scheme for an improved detection of the soliton's density distribution.

%**********************************************
% Concepts
%**********************************************
In addition to studying the soliton’s density profile along the lattice direction, we found it important to also include its radial profile and three-body loss in our models. Although not limiting, three-body loss is non-negligible due to the increased density arising from lattice confinement and attractive interactions. To capture the soliton’s full dynamical behavior, we numerically solved the three-dimensional Gross-Pitaevskii equation (3D-GPE) with an added quintic loss term \cite{saito2002mean, adhikari2002mean, kagan1998collapse}. However, we start by analyzing the system with a variational approach based on a Gaussian ansatz \cite{salasnich2007} to provide initial insight into the soliton’s stability and the underlying physical mechanisms.

\begin{figure}[t]
\includegraphics[width=\columnwidth]{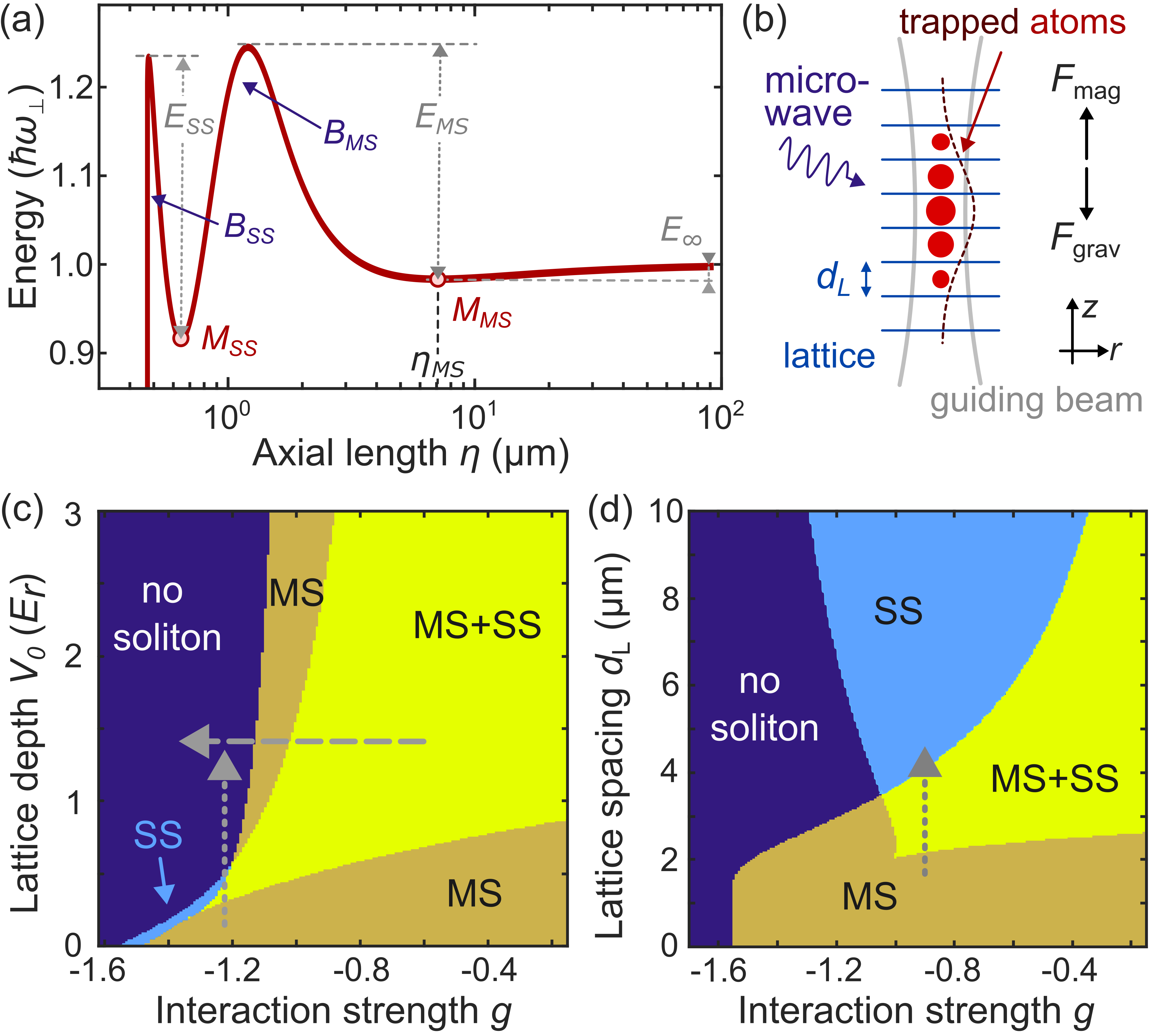}
\caption{Experimental setup and stability diagrams. (a) Energy $E(\eta)$ for a Gaussian wave packet with $V_0=1.1\,E_r$, $a_s=-6.2\,a_0$, $d_\text{L}=2\,\upmu$m. Single-site (SS) and multi-site (MS) solitons are stable at minima $M_{SS}$ and $M_{MS}$ with barriers $B_{SS}$ and $B_{MS}$. (b) Sketch of experimental setup. (c) Stable regions of SS and MS solitons for varying parameters $g$ and $V_0$, with $N=1800$, $\omega_{\perp}=2\pi\times30$\,Hz, $d_\text{L}=3.2\,\upmu$m. No solitons exist in dark blue regions. (d) Stable regions for varying $d_\text{L}$, same parameters as (c) with constant $V_0=1.3\,E_r$ set at $d_\text{L}=3.2\,\upmu$m.}\label{Fig:setup}
\end{figure}

Within this model, the energy of a Gaussian wave packet with axial length $\eta$ and radial width $\sigma$ is given by 
\begin{multline} E = \frac{1}{2}\left( \frac{1}{2\eta^2}+ \frac{1}{\sigma^2} + \sigma^2 \right) + \frac{g}{2\sqrt{2\pi}}\frac{1}{\sigma^2 \eta} - V_0 \exp(-k_\text{L}^2 \eta^2).\\ \label{Eq:energy}
\end{multline}
Here, $\eta,\sigma$ are in units of the radial harmonic oscillator length $a_\perp=\sqrt{\hbar/m\omega_{\perp}}$, and $E$ is in units of $\hbar\omegaP$, where $\omega_{\perp}$ is the radial trap frequency and $m$ is the atomic mass. The first term in Eq.\,(\ref{Eq:energy}) provides the kinetic energy of the soliton, while the second term describes the interaction energy using the interaction strength $g=2 a_s N/a_{\perp}$, where $a_s$ is the s-wave scattering length and $N$ the total atom number. The third term contains the lattice contribution, with lattice depth $V_0$ in units of $\hbar\omegaP$ and wave number $k_L=\pi/d_\text{L}$. For a simplified illustration [Fig.\,\ref{Fig:setup}(a)], we determined the value $\sigma_\text{min}$ that minimizes $E(\eta,\sigma)$ for each value of $\eta$ \cite{carr2002,parker2007}. The resulting energy $E(\eta)=E(\eta,\sigma_\text{min}(\eta))$ shows two minima where stable single-site and multi-site solitons form [$M_{SS}$ and $M_{MS}$ in Fig.\,\ref{Fig:setup}(a)]. Collapse towards smaller axial length $\eta$ is prevented by two barriers $B_{SS}$ and $B_{MS}$.

Without a lattice potential, there is only one barrier with a single critical interaction strength, $g_c$ \cite{parker2007}, beyond which the barrier disappears and the wave packet collapses. The value of $g_c$ depends on geometry and confinement, and various methods have been used for its predictions, e.g., numerically solving the full 3D Gross-Pitaevskii equation \cite{carr2002,ruprecht1995} with a variational approach \cite{perez-garcia1997,carr2002}, or using the nonpolynomial Gross-Pitaevskii equation \cite{salasnich2002}. With a lattice potential, the barrier heights depend also on lattice depth and spacing, and $g_c$ is replaced by surfaces in the ($g, V_0, d_\text{L}$)-parameter space that indicate the disappearance of the barriers. 

Patches in Figs.\,\ref{Fig:setup}(c) and \ref{Fig:setup}(d) represent stable regions with non-zero energy barriers for parameters ($g, V_0$) and ($g, d_\text{L}$). The critical interaction strength is given by the boundary that separates parameter regions without solitons (dark blue patches) and with solitons (``SS'', ``MS''). The interplay between $V_0$, $d_\text{L}$, and $g$, and the barrier heights $E_{SS}$ and $E_{MS}$ is not straightforward. For instance, decreasing $g$ at a fixed lattice depth [dashed horizontal arrow in Fig.~\ref{Fig:setup}(c)], lowers the barriers due to strong attractive interactions and leads to the eventual collapse, first of the single-site soliton followed by the multi-site soliton. Conversely, when the interaction strength is held constant [dotted vertical arrow in Fig.\,\ref{Fig:setup}(c)], the multi-site soliton can already exist at shallow lattice depths, whereas a larger value of $V_0$ is required to form the energy minimum $M_{SS}$ that supports the single-site soliton. A further increase of  $V_0$ eventually eliminates both barriers. Both types of solitons connect to bright 1D solitons without a lattice, either in the limit of vanishing lattice depth or in the limit of large lattice spacing for single-site solitons \cite{SuppMat}.

%***********************
% Experimental setup
%***********************
In our experiment, we created a magnetically levitated Bose-Einstein condensate (BEC) of $N\approx1.3\times10^5$ cesium atoms in a crossed-beam dipole trap at a wavelength of 1064\,nm \cite{kraemer2004, dicarli2019a}. A broad magnetic Feshbach resonance in the $F=3,m_F=3$ state with a zero-crossing at 17.1\,G allowed us to tune interactions \cite{gustavsson2008,berninger2013}. To reduce the atom number, we lowered the levitation gradient over three seconds ($N\approx 30,000$ atoms) before transferring the condensate into our accordion lattice at 780\,nm. All but a few central sites were then selectively cleared using a combination of microwave transfer and resonant light ($N\lesssim3,000$ atoms) [Fig.\,\ref{Fig:setup}(b)]~\cite{peaudecerf2019b}. During the transfer, we set $d_\text{L}=3.2(2)\,\upmu$m and $V_0\approx100\,E_r$, to simplify the spatial site selection in the levitation gradient. $V_0$ is always given in recoil energy, $E_r=(\hbar \pi/d_\text{L})^2/(2m)$, where $d_\text{L}$ is the lattice spacing specific to each measurement. While we can remove 95\% of the atoms per site without affecting neighbors, here we increased the removal to close to 100\% at the cost of ~5\% loss in neighboring sites (for further details see \cite{SuppMat, cruickshank2025}).

To prepare the initial density profile of the wave packet before the interaction quench, we added a dipole trap with frequency $\omega_z$ [dashed line Fig.\,\ref{Fig:setup}(b)], adjusted both $d_\text{L}$ and $V_0$ to their final values, and tuned $a_s$ to approximately $+20\,a_0$ in 400\,ms, where $a_0$ is Bohr's radius. An additional waiting period of 200\,ms ensured phase coherence between the sites, which we verified through free expansion measurements. Finally, we created the solitons by quenching $a_s$ to negative values and by removing the longitudinal trapping potential within 2\,ms. After an evolution time $t$, we used a magnification scheme to analyze the density distribution of the wave packet with absorption imaging \cite{alassam2010}. The lattice depth $V_0$ was increased to approximately $100\,E_r$, effectively freezing the atom distribution within the sites, followed by a slow expansion of $d_\text{L}$ to $20(1)\,\upmu$m over a period of 400\,ms. 

\begin{figure}[t]
\includegraphics[width=\columnwidth]{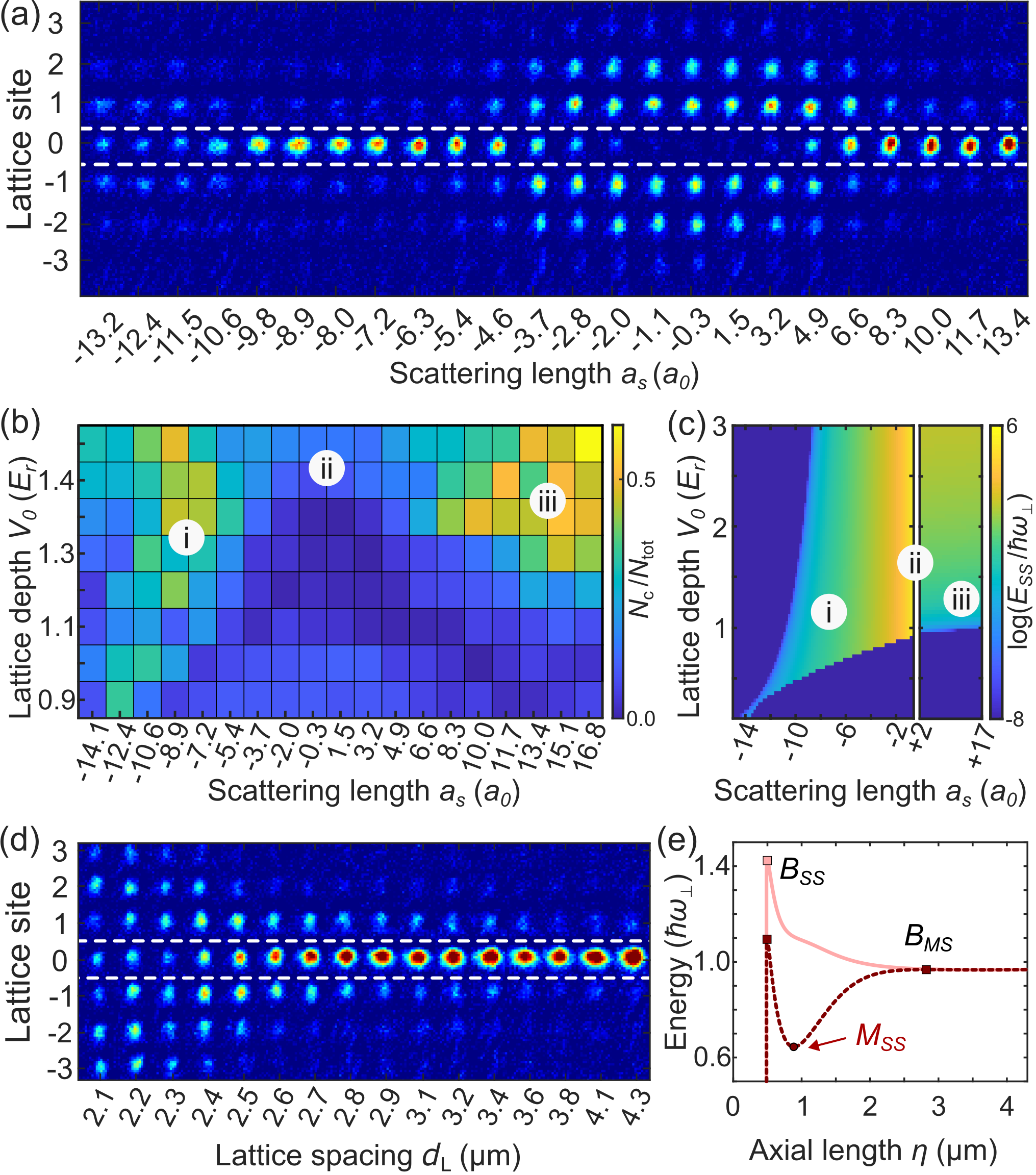} 
\caption{Stability of single-site solitons. (a) Measured density distribution after a quench of $a_s$ and $t=100$\,ms hold time with $d_\text{L}=3.2(2)\,\upmu$m, $V_0=1.3(1)\,E_r$, $\omega_\perp = 2\pi\times 40(1)$\,Hz, $N\approx1800$. White lines mark atoms in the central site. (b) Measured relative central-site atom number $N_c/N_\text{tot}$ vs. $a_s$ and $V_0$ with same parameters as (a). (c) Energy $E_{SS}$ of the barrier $B_{SS}$; (i-iii) indicate regions of varying stability in (b,c). See \cite{SuppMat} for the definition of $E_{SS}$ for $a_s>0\,a_0$. (d) Density distribution for varying $d_\text{L}$ after 100\,ms with $a=-6.4\,a_0$, $N\approx1800$, constant $V_0=1.3(1)\,E_r$ set at $d_\text{L}=3.2(2)\upmu$m. (e) Calculated $E(\eta)$ for (d) with $d_\text{L}=3.5\,\upmu$m (dotted line) and  $d_\text{L}=2.0\,\upmu$m (solid line). Measured data is averaged over typically seven repetitions.}\label{Fig:singlesite}
\end{figure}

In a first measurement, we demonstrated the existence and properties of single-site solitons. After preparing approximately $1800$ atoms at a single site, we quenched $a_s$ and measured the density profile and the atom number per lattice site after a hold time of 100\,ms. Absorption images of the density profile show a strong dependence on $a_s$ [Fig.\,\ref{Fig:singlesite}(a)]. For $a_s\approx-8\,a_0$, the wave packet remained localized at the central lattice site, which indicates the formation of a single-site soliton. Except for some initial shedding of atoms, we found this soliton to be stable for a hold time up to 2\,s \cite{SuppMat}. For stronger attractive interaction, $a_s<-10\,a_0$, the soliton collapsed, and the remaining atoms spread along the lattice direction. Weak attractive and repulsive interactions, $-5\,a_0<a_s<+5\,a_0$, resulted in the dispersion of the wave packet, with a minimum at the central lattice site after the given hold time, while larger scattering lengths $a_s>+7\,a_0$ lead again to the localization of the wave packet. In the two-particle limit, this localized state corresponds to repulsively bound pairs~\cite{winkler2006}, whereas in the context of two lattice sites and Josephson oscillations, it is associated with macroscopic quantum self-trapping~\cite{smerzi1997,albiez2005a}.

We extended the study to different lattice depths and determined the relative atom number in the central site, $N_c/N$, as a measure of the system stability [Fig.\,\ref{Fig:singlesite}(b)]. The data reveals the three regimes: (i) a stable single-site soliton, (ii) a free dispersion of the wave packet close to $0\,a_0$, and (iii) the self-trapping for repulsive interaction and sufficient lattice depth. The regimes can be explained by the height of the barriers $B_{SS}$ and $B_{MS}$. For comparison, Fig.\,\ref{Fig:singlesite}(c) shows the energy $E_{SS}$, which is the height of barrier $B_{SS}$ [see Fig.\,\ref{Fig:setup}(a) and \cite{SuppMat}]. Large values of $E_\text{SS}$ align well with the experimental data in Fig.\,\ref{Fig:singlesite}(c), accurately predicting the stable regions (i) and (iii). However, $E_{SS}$ does not capture the evolution of the wave packet close to zero scattering length in region (ii), where the wave packet spreads. While collapse is prevented by barrier $B_{SS}$, spreading is inhibited by the barriers at larger values of $\eta$ [Fig.\,\ref{Fig:setup}(a)].

To investigate the effect of the lattice spacing on the stability of the solitons, we varied $d_\text{L}$ while keeping $V_0$ and $a_s$ constant [Fig.\,\ref{Fig:singlesite}(d)]. Absorption images taken after a hold time of $100$\,ms show a spreading of the wave packet for $d_\text{L}\lesssim 2.5\,\upmu$m [Fig.\,\ref{Fig:singlesite}(d)]. Our calculations of $E(\eta)$ indicate that as $d_\text{L}$ decreases, the minimum $M_{SS}$ disappears, while the barrier $B_{SS}$ persists [Fig.\,\ref{Fig:singlesite}(e)]. Consequently, the observed spreading after the interaction quench is not due to a collapse, as observed in Fig.\,\ref{Fig:singlesite}(a), but rather due to the absence of an energy minimum. The calculated minimum $M_{SS}$ vanishes for $d_\text{L}\approx 2.2\,\upmu$m, which agrees with our experimental data ($d_\text{L}\approx2.5\,\upmu$m). This measurement also aligns well with the stability diagram in Fig.\,\ref{Fig:setup}(d) (dotted arrow). 

%**********************************************
% Multisite soliton
%**********************************************
\begin{figure}[b]
\includegraphics[width=\columnwidth]{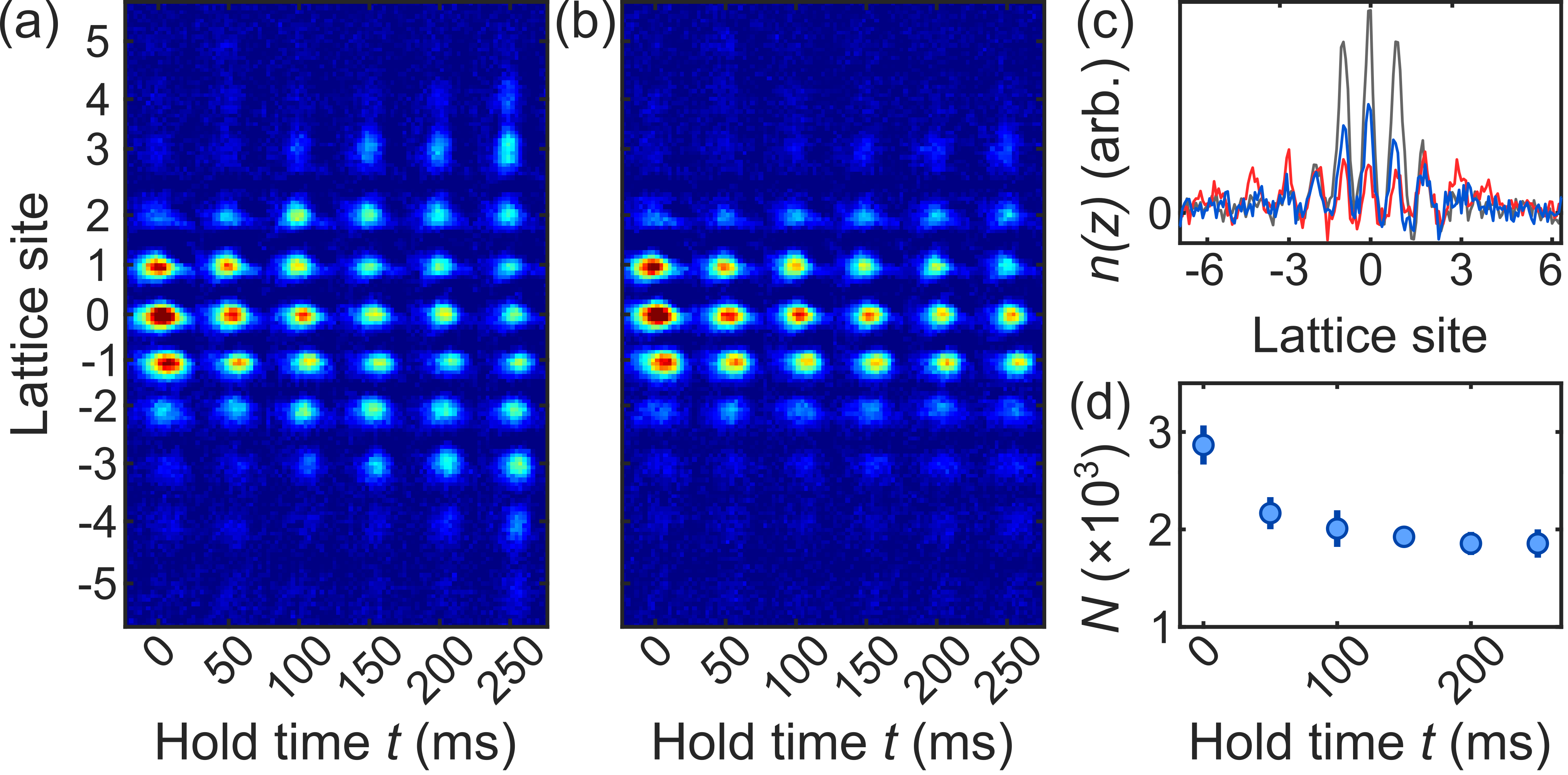}
\caption{Stability of a multi-site wave packet. (a,b) Time evolution of a wave packet after a quench of $a_s$, averaged over ten repetitions with $V_0=1.3\,E_r$, $d_\text{L}=2.6\,\upmu$m, $\omega_\perp=2\pi\times 25\,$Hz, $\omega_z=2\pi\times25$\,Hz, $N\approx2900$. (a) The wave packet disperses for a quench to $a_s=+2.0\,a_0$ and (b) mostly retains its overall shape for $-5.7\,a_0$. Site occupation numbers for both data sets are provided in Ref.\,\cite{SuppMat}. (c) Density profiles of the wave packet immediately after the quench (gray), and after $250$\,ms for $+2.0\,a_0$ (red) and $-5.7\,a_0$ (blue). (d) Atom number for data in (b), error bars denote the standard deviation. A comparison of the data in (c,d) with a 3D-GPE simulation is provided in Ref.\,\cite{SuppMat}.}\label{Fig:multisite}
\end{figure}

In a second measurement, we investigated the stability of multi-site wave packets. To prepare the initial state, we adjusted the microwave transfer to remove all but the atoms in three adjacent lattice sites. During the subsequent waiting period, this density profile evolved toward a Gaussian envelope spanning 3–5 lattice sites, determined by the trapping frequency $\omega_z$. After the quench, we observed stronger density fluctuations compared to single-site solitons, showing in some cases a splitting of the soliton with moving fractions. Quantum fluctuations have been suggested as a possible cause of this fragmentation \cite{dabrowska-wuster2009,streltsov2011,martin2012}. However, here, we attributed it to technical noise and the low binding energy of multi-site solitons, $E_\infty$ [Fig.\,\ref{Fig:setup}(a)].

We studied the time evolution of the wave packet over 250\,ms following the quench. For scattering lengths near zero [Fig.\,\ref{Fig:multisite}(a)], the wave packet shows dispersion, whereas for $a_s=-5.7\,a_0$ [Fig.\,\ref{Fig:multisite}(b)], it remains mostly localized. We attribute this localization to the formation of a multi-site soliton-like state. The density profiles show little change in site occupations between $t=100$\,ms and $250\,$ms, indicating a stable configuration. However, during the initial tens of milliseconds, we observe a loss of atoms in the three central sites. We attribute this initial depletion to three-body loss, which predominantly occurs in high-density regions where attractive interactions enhance the local density \cite{weber2003}. The subsequent atom loss is strongly reduced, and we speculate that the wave packet gradually adjusts its radial and axial sizes in response to the slowly decreasing atom number.

\begin{figure}[t]
\includegraphics[width=\columnwidth]{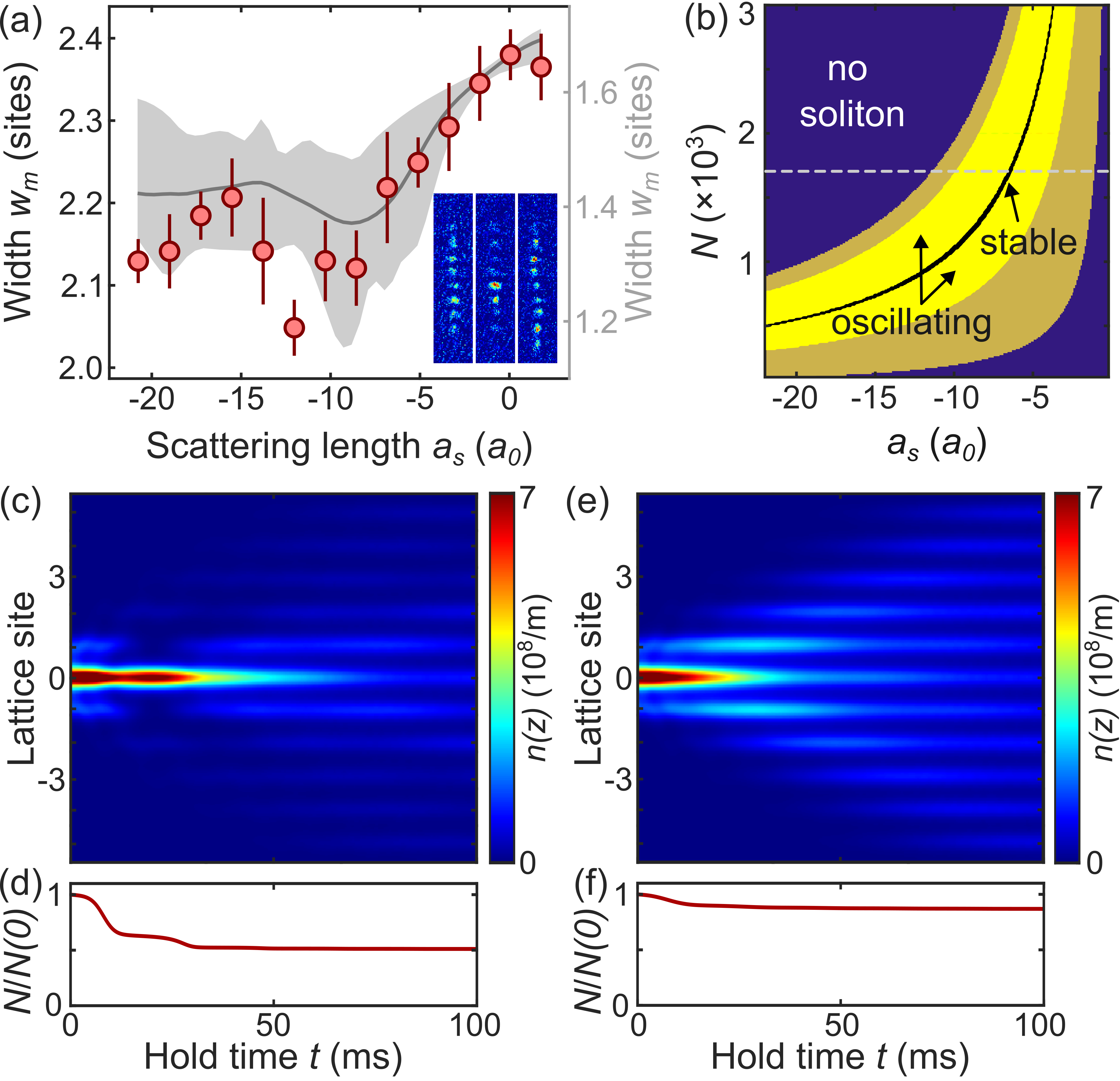}
\caption{Collapse of a multi-site wave packet. (a) Width $w_m$ of the wave packet at $t=150\,$ms after quenching to different values of $a_s$, with $V_0=1.4\,E_r$, $d_\text{L}=2.6\,\upmu$m, $\omega_\perp=2\pi\times 30\,$Hz, $N\approx1700$. The gray patch shows the variation in $w_m$, calculated using the 3D-GPE, resulting from uncertainties in the three-body loss coefficient $L_3$ and $N$. The line is an average of the calculations \cite{SuppMat}. (Inset, left to right) Typical images of the density profiles after collapse ($a_s=-17\,a_0$), shrinking towards the central site ($a_s=-10\,a_0$) and expanding wave packet ($a_s=0\,a_0$). (b) Stability regions calculated using Eq.\,(\ref{Eq:energy}) with an existing minimum $M_{MS}$ (brown) and without (blue), with breathing oscillations for $E(\eta)<E_\infty$ and $E_{MS}$ (yellow), and with stable multi-site solitons for $E(\eta_0)\approx E_{MS}$ (black). (c),(d) The calculated time evolution of the density distribution and relative atom number show collapse followed by expansion for $a_s=-9.5\,a_0$, and (e),(f) dispersion for $a_s = -1.7\,a_0$. Calculations use $L_3=5\times 10^{-39}\,$m$^6$s$^{-1}$ and other parameters as in (a). }\label{Fig:collapse}
\end{figure}

The 1D density profiles [Fig.\,\ref{Fig:multisite}(c)] and the measured total atom number [Fig.\,\ref{Fig:multisite}(d)] support this interpretation. Comparing the initial 1D density profile [gray line in Fig.\,\ref{Fig:multisite}(c)] with the profile after 250\,ms reveals a reduction in atom number at the three central sites for the soliton-like wave packet (blue line), while the overall profile remains localized. In contrast, for near-zero interactions (red line), the wave packet undergoes significant spreading, with increased occupation of the outermost sites. To quantify the spreading, we calculated the relative site occupations $N_j/N$, and extracted the wave packet width \cite{SuppMat}. The non-interacting and soliton-like wave packets exhibit linear dispersion velocities of 12 and 7\,sites/s, respectively. For the soliton-like wave packet, this apparent spreading mainly reflects the flattening of the density profile rather than significant mass transport. We also found good agreement of the density profiles and atom numbers in Figs.\,\ref{Fig:multisite}(c) and \ref{Fig:multisite}(d) with simulations of the 3D-GPE \cite{SuppMat}.

Finally, we quantitatively analyzed the collapse by measuring the wave packets' density profiles at $t = 150$\,ms over a broad range of scattering lengths [Fig.\,\ref{Fig:collapse}(a)].
To account for varying density distributions, we calculated the second-moment width $w_m$ of the site occupations, defined as
\begin{align*}
    w_m^2 = \frac{1}{N}\sum_j N_j (z_j-\bar{z})^2, \qquad \text{with} \quad j=-4,...,4.
\end{align*}
Here, $z_j$ is the position of the $j$th lattice site and $\bar{z}$ is the center-of-mass position. The value of $w_m$ indicates the varying stability of the wave packet depending on $a_s$. It spreads for $a_s\approx 0\,a_0$, shrinks towards the central site for $a_s\approx -10\,a_0$, and spreads after collapse for strong attractive interactions $a_s< -13\,a_0$. Single absorption images illustrate the spreading and shrinking of the wave packet in the different regions [inset in Fig.\,\ref{Fig:collapse}(a)].

The variational approach used in Eq.\,(\ref{Eq:energy}) provides a simple model for predicting the evolution of the wave packet after quenching to scattering length $a_s$. Within the model, stability is achieved when the initial parameters of the wave packet, $N(0)$ and $\eta_0$, closely match those of a multi-site soliton with length $\eta_{MS}$. In our experimental protocol, $N(0)$ and $\eta_0$ are set during preparation, while only $a_s$ can be varied. A soliton is created by quenching the scattering length to $a^*_s$ with $\eta_0=\eta_{MS}(N(0),a_s^*)$. For other quench values close to $a_s^*$, the wave packet is expected to exhibit small breathing oscillations \cite{salasnich2007}, unless its initial energy $E(\eta_0)$ exceeds one of the barrier energies, $E_\infty$ or $E_{MS}$, leading to dispersion or collapse. Calculating the barrier energy barriers $E_\infty$ and $E_{MS}$ using Eq.\,(\ref{Eq:energy}) allows us to predict the stability regions. The brown patch in Fig.\,\ref{Fig:collapse}(b) marks where a minimum $M_{MS}$ exists. Stable solitons form only along the black line, while breathing oscillations occur within the yellow regions. Assuming a fixed atom number further constrains the choice of $a_s$ to lie on the dashed line, though in practice $N$ decreases due to three-body loss.

To capture the full evolution of the wave packet beyond this simple model, we numerically simulated the dynamics of the multi-site soliton using a modified 3D Gross-Pitaevskii equation (GPE) with a quintic term that accounts for three-body loss \cite{saito2002mean, adhikari2002mean, SuppMat}. The simulations show two distinct dynamical regimes. In the first regime, corresponding to large negative values of $a_s$, the wave packet begins to collapse, leading to an increase in local density at the central site [Fig.\,\ref{Fig:collapse}(c)]. However, a further shrinking of the wave packet is suppressed by the enhanced loss and a rapid shedding of atoms [Fig.\,\ref{Fig:collapse}(d)]. The second regime, which occurs for less negative values of $a_s$, is marked by a slow dispersion of the matter wave, and has lower and more gradual atom loss [Fig.\,\ref{Fig:collapse}(e) and (f)]. 

The simulation agrees well with the observed shrinking in $w_m$ during the collapse process. However, it is sensitive to the precise values of atom number and the three-body loss coefficient \cite{weber2003,kraemer2006}, resulting in an uncertainty of the predicted dynamics [gray patch in Fig.\,\ref{Fig:collapse}(a)] \cite{SuppMat}. In addition, imaging noise in the experimental data increases $w_m$, leading to an offset and reduced contrast compared to the simulation.
While the observed atom loss was sufficiently low to permit the formation and investigation of lattice solitons, its inclusion in our simulation was still essential to reproduce our observations. Interestingly, at strong attractive interactions, the loss helped to suppress collapse and enhanced the stability of the system.

%**********************************************
% Multisite soliton
%**********************************************

In conclusion, we have demonstrated the existence and stability of both single-site and multi-site solitons that extend over varying numbers of lattice sites. Using an accordion lattice with adjustable lattice spacing, we examined their properties across various lattice depths and spacings, and compared our findings with theoretical predictions. A variational model based on a Gaussian approximation for the solitons was used to identify stable parameter regions, while numerical simulations of the 3D-GPE with a three-body loss term captured the solitons' time evolution. We found both types of solitons to be stable for hundreds of milliseconds, allowing ample time for further studies. 

Our results pave the way for exploring a multitude of nonlinear matter-wave excitations in optical lattices, such as lattice breathers \cite{flach2008} and discrete solitons in deep lattice potentials, described by the discrete nonlinear Schrödinger equation \cite{gligoric2009,maluckov2008}. For example, our approach allows investigating the Peierls-Nabarro barrier \cite{ahufinger2004a}, probing 2D solitons \cite{baizakov2003}, and experimentally accessing the dynamical phase diagram \cite{trombettoni2001a,franzosi2011}, which predicts the emergence of breathers and solitons as a function of quasimomentum. Our results advance the understanding of nonlinear wave dynamics in structured media and open new avenues for technological applications, e.g., in matter-wave interferometry \cite{mcdonald2014a, helm2015, polo2021}, precision sensing \cite{naldesi2022}, and the controlled transport of atomic wave packets for quantum information processing \cite{tutunnikov2023, sedov2014}.

\vspace{1ex}

The data used in this publication are openly available at the University of Strathclyde Knowledge Base \cite{DataBase}.
\vspace{1ex}

We acknowledge support by the EPSRC through a New Investigator Grant (EP/T027789/1), the Programme Grant ``Quantum Advantage in Quantitative Quantum Simulation'' (EP/Y01510X/1), and the Quantum Technology Hub in Quantum Computing and Simulation (EP/T001062/1). TH acknowledges funding from the European Research Council (ERC Starting Grant ``FOrbQ'', 101165353). FL and LS are supported by the ``Iniziativa Specifica Quantum'' of INFN and by the project ``Frontiere Quantistiche'' (Dipartimenti di Eccellenza) of the Italian Ministry of University and Research (MUR). LS is supported by the European Union through the European Quantum Flagship Project ``PASQuanS2'', the National Center for HPC, the Big Data and Quantum Computing [Spoke 10: Quantum Computing]. LS also acknowledges funding by the PRIN project ``Quantum Atomic Mixtures: Droplets, Topological Structures, and Vortices'' of MUR. 
\vspace{3ex}

\onecolumngrid
\vspace{5ex}
\begin{center}
\begin{tikzpicture}
  \draw[line width=0.5pt] (-4,0) -- (4,0);
  \draw[line width=0.75pt] (-3,0) -- (3,0);
  \draw[line width=1.0pt] (-2,0) -- (2,0);
  \draw[line width=1.25pt] (-1,0) -- (1,0);
\end{tikzpicture}
\end{center}
\vspace{5ex}

\twocolumngrid

% S figures
\renewcommand{\thefigure}{S\arabic{figure}}
\setcounter{figure}{0}

\begin{center}
\textbf{\Large Supplemental material}
\end{center}

\section{Experimental setup and data analysis}

{\bf Accordion lattice.} 
To achieve the variability in the lattice spacing $d_\text{L}$, we make use of an optical accordion lattice~\cite{fallani2005, williams2008, alassam2010}. Details of our setup are based on Ref.\,\cite{li2008, ville2017}. Using an acousto-optical deflector and a combination of polarizing beam-splitter cubes, we change the angle $\theta$ between the interfering laser beams that create the lattice potential. The resulting lattice spacing depends on $\theta$ with
\begin{align*}
    d_\text{L} = \frac{\lambda}{2\sin(\theta)}.
\end{align*}  
By varying the driving frequency of the acousto-optical deflector, we can tune $\theta$ with a bandwidth of several tens of kilohertz or, as required in this work, implement slow, controlled ramps of $d_\text{L}$ to minimize excitations of the atoms. Using this setup, we achieve lattice spacings ranging from $1.5\,\upmu$m to $40\,\upmu$m.
\vspace{1ex}
\begin{figure}[h]
\includegraphics[width=0.9\columnwidth]{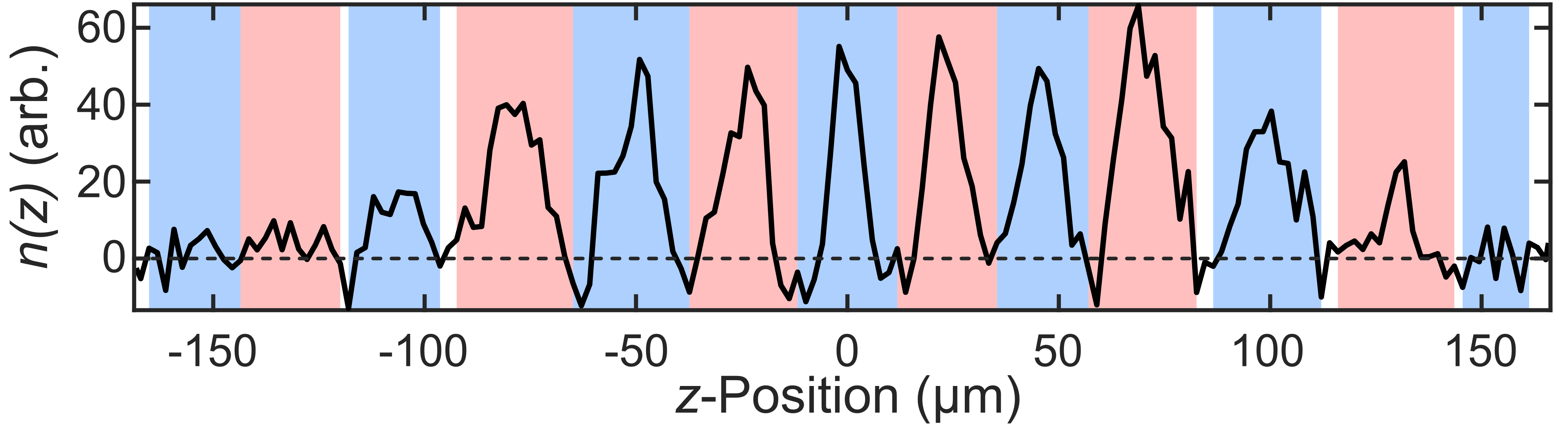}  
\caption{Measurement of the atom number per lattice site. Density profile of a wave packet after a hold time of $250$\,ms, with $V_0=1.3\,E_r$, $d_\text{L}=2.6\,\upmu$m, $\omega_\perp=2\pi\times 25\,$Hz, $N\approx2200$, $a_s=+2.0\,a_0$. Patches denote regions used for the calculation of occupation numbers, $N_j$, at the respective sites.\label{Fig:S1}}
\end{figure}
{\bf Atom number per lattice site.} 
We determined the atom number $N_j$ at each lattice site $j$ from absorption images by first computing the one-dimensional density profile $n(z)$ and then summing the signal within individual lattice sites [Fig.\,\ref{Fig:S1}]. In the images, we observed slight variations in the spacing between neighboring density peaks, caused by our magnification scheme. Specifically, the increase in lattice spacing $d_\text{L}$ during the magnification process introduced small-amplitude oscillations of the atoms within each site, leading to nonuniform peak spacing. To account for these variations, we avoided direct integration over fixed-width regions, but employed a minimum-finding algorithm to dynamically set the integration boundaries for each site. The patches in Fig.\,\ref{Fig:S1} illustrate these boundaries. To avoid nonphysical atom number estimates, we excluded negative values in the density profile that arise from imaging noise and weak diffraction artifacts.

\vspace{1ex}

\begin{figure}[h]
\includegraphics[width=0.95\columnwidth]{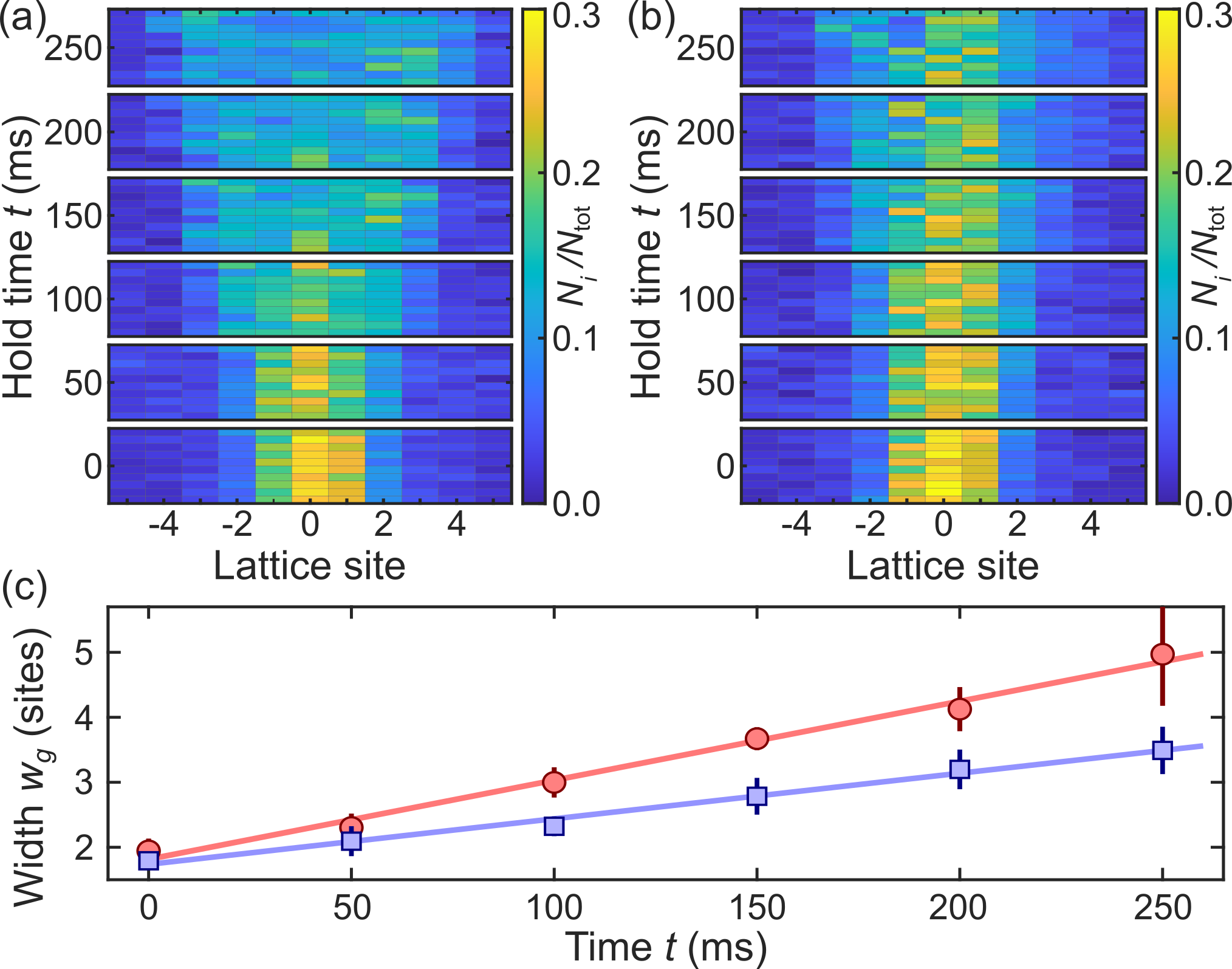}  
\caption{Multi-site solitons. Occupation numbers $N_j$ for the time evolution with a scattering length (a) $a_s=+2.0\,a_0$, and (b) $a_s=-5.7\,a_0$. Parameters $V_0=1.3\,E_r$, $d_\text{L}=2.6\,\upmu$m, $\omega_\perp=2\pi\times 25\,$Hz, $N\approx2200$. The panels group measurements of equal hold time, each with ten repetitions. The same raw data is used as for the averaged images in Fig.\,3 of the main text. (c) Widths of the wave packets for the data in (a) - red circles, and (b) - blue squares. Lines with corresponding colors indicate linear fits to determine $\Delta w_g/\Delta t$. \label{Fig:S2}}
\end{figure}

{\bf Shot-to-shot fluctations of $N_j$.} 
Our measured density profiles show weak shot-to-shot fluctuations of $N_j$ which we attribute to small variations of $N$ and the magnetic field. To illustrate these fluctuations, we show the time evolution of the relative occupation numbers $N_j(t)/N(t)$ for the data in Fig.\,3(a) and (b) of the main text [Fig.\,\ref{Fig:S2}(a) and (b)]. Each horizontal panel groups ten repetitions for the same hold time. As the hold time increases, density fluctuations become more pronounced, as in the panel at $t=250$\,ms in Fig.\,\ref{Fig:S2}(b). We attribute the observed asymmetry in the density profiles at longer times to weak residual forces of magnetic field gradients, which may cause fragmentation and displacement of the wave packet during evolution.
\vspace{1ex}

{\bf Dispersion velocities.}
We characterized the width and spreading of the wave packets using a Gaussian envelope. The distribution of occupation numbers $N_j$ was fitted with the function $$n(z_j) = a\exp(-(z_j-z_0)^2/w_g^2),$$
where $a$, $z_0$, and $w_g$ are fitting parameters. The extracted widths $w_g(t)$ show a linear increase over time. To quantify this spreading, we performed linear fits [lines in Fig.\,\ref{Fig:S2}(c)], resulting in dispersion velocities of $\Delta w_g/\Delta t= 12\,$sites/s and $7$ sites/s for the data in Figs.\,\ref{Fig:S2}(a) and (b), respectively. The observed dispersion of the soliton is primarily caused by a flattening of its density profile due to particle loss, creating the appearance of spreading despite little mass transport between sites (see main document).
\vspace{1ex}

\begin{figure}[h]
\includegraphics[width=0.9\columnwidth]{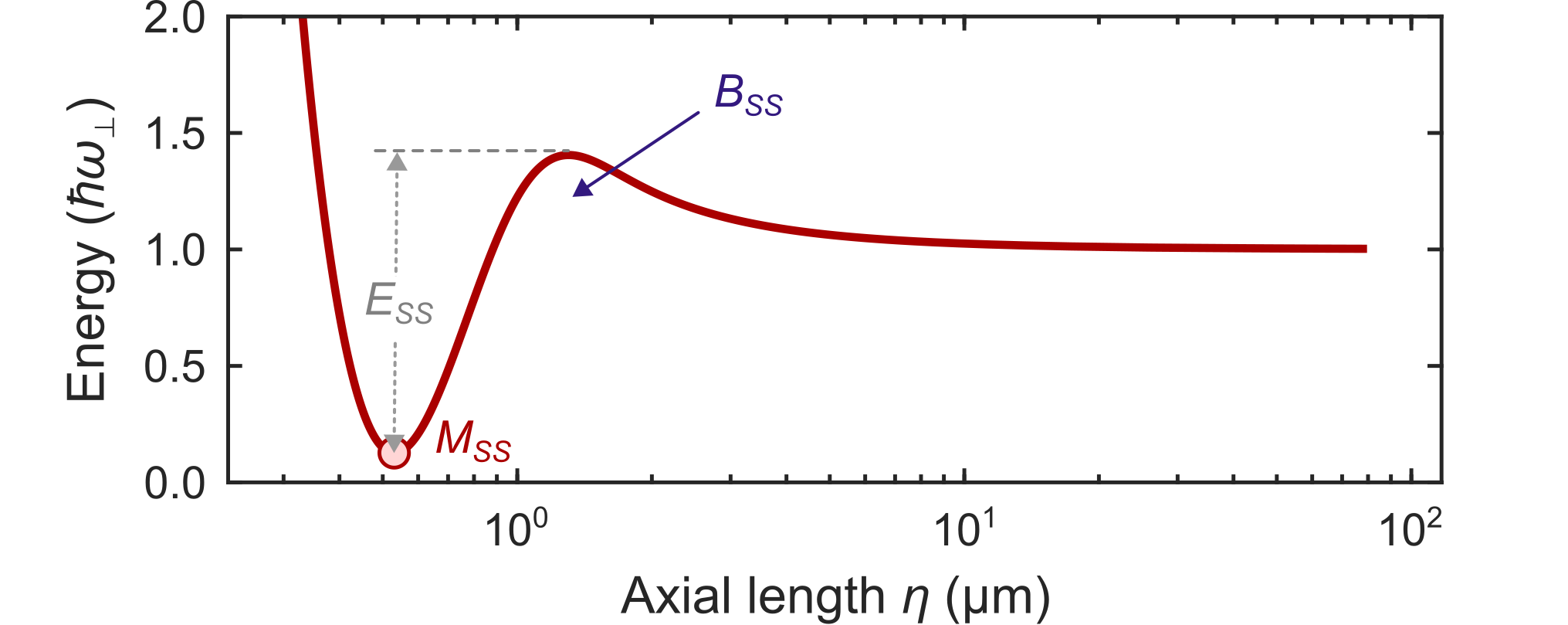}  
\caption{Energy minimum for repulsive interaction. Energy $E(\eta)$ in Eq.\,(1) for a Gaussian wave packet with axial length $\eta$ and lattice depth $V_0=2.2\,E_r$, scattering length $a_s=+5.0\,a_0$, and lattice spacing $d_\text{L}=2.0\,\upmu$m. A minimum $M_{SS}$ forms with an energy barrier $B_{SS}$ that prevents spreading of the wave packet.\label{Fig:S3}}
\end{figure}

{\bf Repulsive interaction.} 
To further illustrate the localization of a single-site wave packet with repulsive interaction, as observed in Fig.\,2(a), we calculated the energy profile $E(\eta)=E(\eta,\sigma_\text{min}(\eta))$ for a wave packet with scattering length $a_s=+5\,a_0$ [Fig.\,\ref{Fig:S3}]. The resulting energy curve~\cite{carr2002,parker2007} shows a single minimum, $M_{SS}$, that allows for the formation of stable wave packets that are localized on a single lattice site. Although repulsive interactions typically lead to spreading, this behavior is suppressed by the energy barrier $B_{SS}$ of height $E_{SS}$, which stabilizes the localized state.

This behavior closely resembles that of a single-site soliton with attractive interaction. However, instead of collapsing, the wave packet here tends to spread due to repulsion. For consistency, we applied the same labels $B_{SS}$ and $E_{SS}$ as used for the single-site soliton. Figure\,2(c) in the main document shows the barrier height $E_{SS}$ for attractive and repulsive interactions. 
\vspace{1ex}

\begin{figure}[h]
\includegraphics[width=0.95\columnwidth]{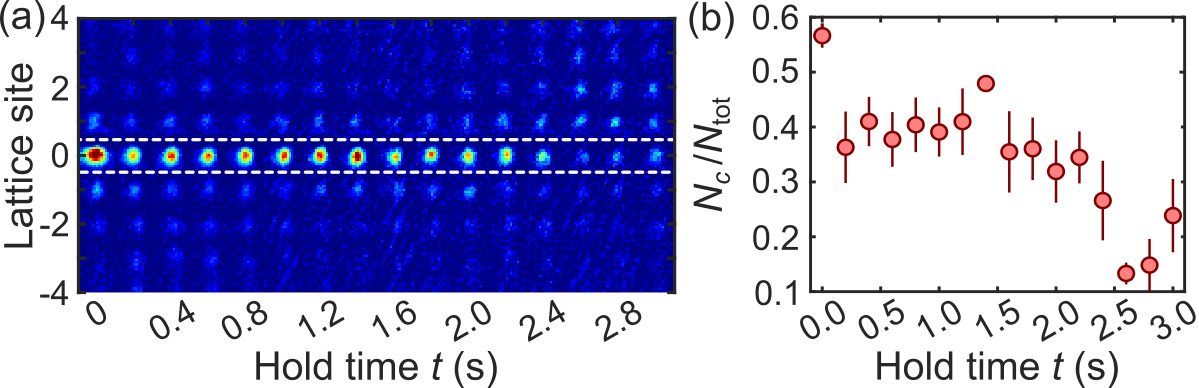}  
\caption{Time evolution of single-site soliton. (a) Average measured density distribution of the wave packet after a quench of the scattering length $a_s$ and for a varying hold time with $d_\text{L}=3.2(2)\,\upmu$m, $V_0=1.3(1)\,E_r$, $a_s=-6\,a_0$, $\omega_\perp = 2\pi\times 40(1)$\,Hz, $N\approx1800$. White lines mark atoms at the central lattice site. (b) Relative atom number $N_c/N$ in the central site for the data in (a). Error bars denote standard errors.\label{Fig:S4} }
\end{figure}

{\bf Lifetime of the single-site soliton.} 
In addition to the measurements in Fig.\,2 of the main text, we determined the lifetime of the single-single site soliton. Averaged absorption images show the density profile of the wave packet after a variable hold time $t$ following the quench [Fig.\,\ref{Fig:S4}(a)]. 

The wave packet remains stable for approximately 2\,s before drifting away from the central site and beginning to spread. This behavior is also reflected in the extracted atom number at the central site $N_c$, normalized to the total atom number within the central 9 sites [Fig.\,\ref{Fig:S4}(b)]. We also observe a rapid drop in atom number within the first $200$\,ms, which we attribute to shedding of atoms and three-body loss after quenching to attractive interaction~\cite{weber2003}.
\vspace{1ex}

{\bf 1D solitons without a lattice potential.} 
The variational model for lattice solitons, shown in Fig.\,1 of the main text, connects well with corresponding models for solitons without a lattice potential. In the limit of vanishing lattice depth, $V_0 \rightarrow 0$, the potential barriers disappear, preventing the formation of lattice solitons for $g<g_c = -1.55$ [Fig.\,1(c)]. A similar limit arises in Fig.\,1(d) for single-site solitons when the lattice depth is held constant and the lattice spacing becomes large, $d_L \rightarrow \infty$, as the trapping frequency at each site scales as $1/d_L$. In this regime, the model predicts single-site solitons persisting down to $g_c = -1.55$. Both critical values are consistent with predictions from variational models without a lattice potential, using a Gaussian ansatz, which yields $g_c = -1.556$ \cite{perez-garcia1997, salasnich2007}.

\vspace{1ex}

{\bf Single-particle effects.}
The long-time evolution of the wave packet in a lattice differs from that of solitons in uniform 1D systems. In both cases, the wave packet is not confined along the axial direction and will be subject to weak residual forces. In the absence of a lattice potential, such forces merely accelerate the soliton and shift its position. However, in the presence of a periodic lattice, they can give rise to Bloch oscillations, which can manifest as position oscillations and breathing motion.

To avoid misinterpreting such single-particle effects as interaction-driven localization or breathing, we consistently compared the evolution with and without interactions throughout this work [see Fig.\,2(a) and Figs.\,3(a) and 3(b)]. This approach allowed us to demonstrate that the observed localization arises from nonlinear, interaction-driven effects.

\section{Numerical simulations}

We performed numerical simulations with the Gross-Pitaevskii equation (GPE) for the condensate wavefunction $\psi$, normalized to the atom number $N$,
\begin{equation}
    i\hbar\frac{\partial}{\partial t} \psi = -\frac{\hbar^2}{2m} \nabla^2\psi + V\psi + g|\psi|^2\psi - i g_5 |\psi|^4\psi \, ,
\end{equation}
where $m = 133u$ is the cesium mass, $g = 4\pi\hbar^2 a_s/m$ is the cubic nonlinearity coefficient due to zero-momentum s-wave scattering with scattering length $a_s$, and the quintic coefficient $g_5$ represent three body losses. The three-body-loss term is given by $g_5=\hbar L_3/2$, with $L_3$ being the three-body loss coefficient, which in the present case is estimated to be $L_3 \sim 10^{-39}$\SI{}{m^6s^{-1}}. The external potential is $V(x, y, z) = \omega_x^2 m \ z^2 / 2 + \omega_\perp^2 m (x^2 + y^2) / 2 + V_0\cos(2k_\text{L}z)$. By including the dissipative term, the normalization of the wavefunction can change with time. 

\begin{figure}[t]
    \centering
    \includegraphics[width=\linewidth]{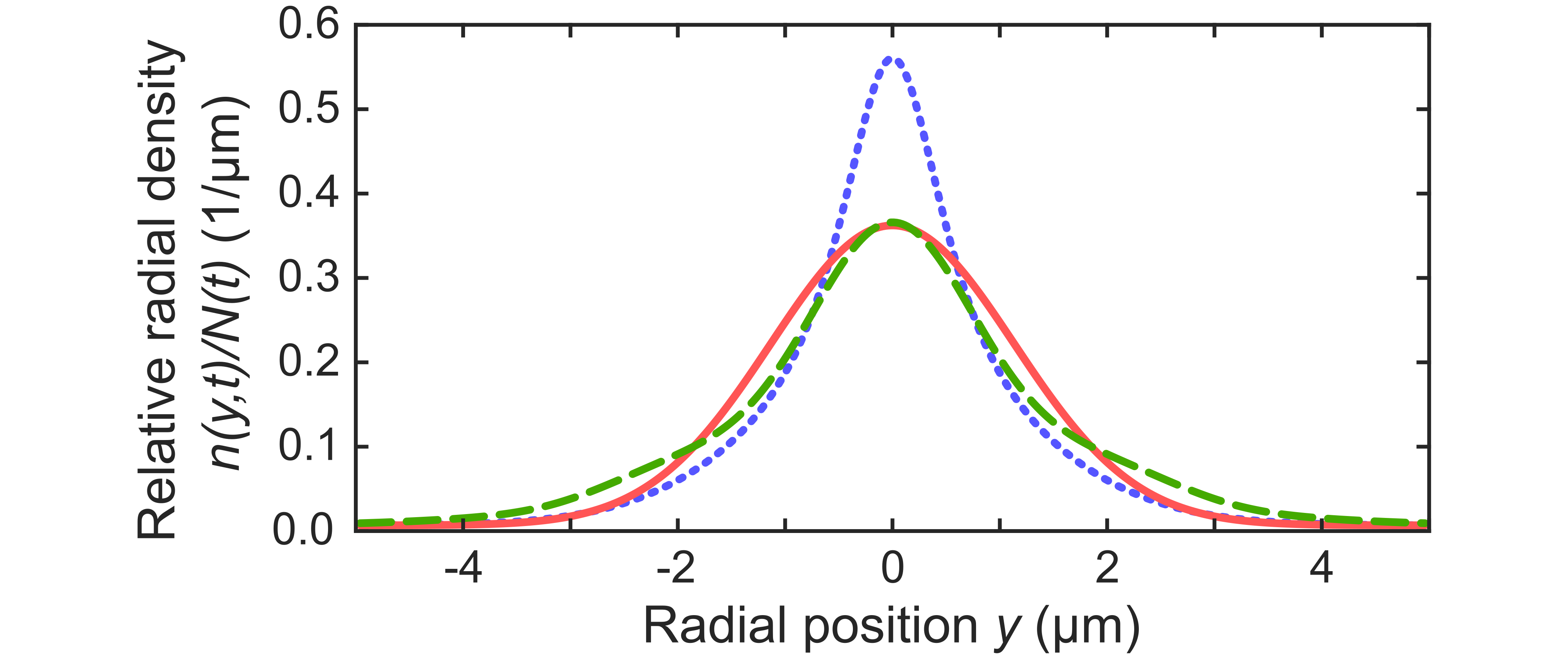}
    \caption{Radial density profile $n(y) = \int dzdx | \psi(x,y,z)|^2$, calculated using the 3D-GPE for a scattering length of $a_s=-18.17\,a_0$. The simulation follows the protocol described in the main text. The profile is shown at two different times: $t=5.13\,$ms (blue dotted line) and $t=10.26\,$ms (green dashed line). The solid red line represents a Gaussian profile for comparison.\label{fig:final_states2} }    
\end{figure}

The 3D-GPE is preferred over other dimensionally-reduced models, such as the Nonpolynomial Schr\"odinger equation (NPSE) and the 1D GPE, because of the special role of the collapse in the experiment. Indeed, the 1D GPE is not sensitive to the collapse, and the NPSE is accurately describing it for stationary solution but in a dynamical evolution it is affected by the vanishing of the transverse width. While being more computationally demanding, the 3D-GPE can describe accurately the transverse dynamics that is crucial near the collapse time. 

We find that the time evolution of the wave packet is highly sensitive to the precise values of $N$ and $L_3$. To indicate the resulting range of possible widths $w_m$ of the wave packet for the measurement in Fig.\,4(a), we vary $L_3$ from $5\times 10^{-39}\,$m$^6$s$^{-1}$ to  $5\times 10^{-38}\,$m$^6$s$^{-1}$ and $N$ from 1200 to 2200 atoms, each in three discrete steps. The gray shaded region in Fig.\,4(a) of the main text shows the envelope defined by the maximal and minimal values of $w_m$, along with the average across all simulation runs (gray line). The variations in the boundaries of the shaded region reflect the fluctuations arising from the different parameter combinations.

To illustrate the radial evolution of the wavefunction, we present two radial projections at different times [Fig.\,\ref{fig:final_states2}] for the collapsing dynamics obtained in the case of $a_s=-18.17 \, a_0$ and $N=1700$, with the initial condition as described in the main text. They are taken slightly before and slightly after the collapse event, respectively in the blue dotted line ($t=5.13$\,ms) and the green dashed line ($t=10.26$\,ms). They are compared to the Gaussian transverse wavefunction obtained solving exactly the ground state in the noninteracting case, which is used as an ansatz for the transverse wavefunction in the 1D GPE. The non-Gaussianity of the transverse distribution suggests the need to utilize the 3D-GPE in analyzing the dynamics near the collapse, even in presence of losses.
\vspace{1ex}

\begin{figure}[b]
    \centering
    \includegraphics[width=0.95\linewidth]{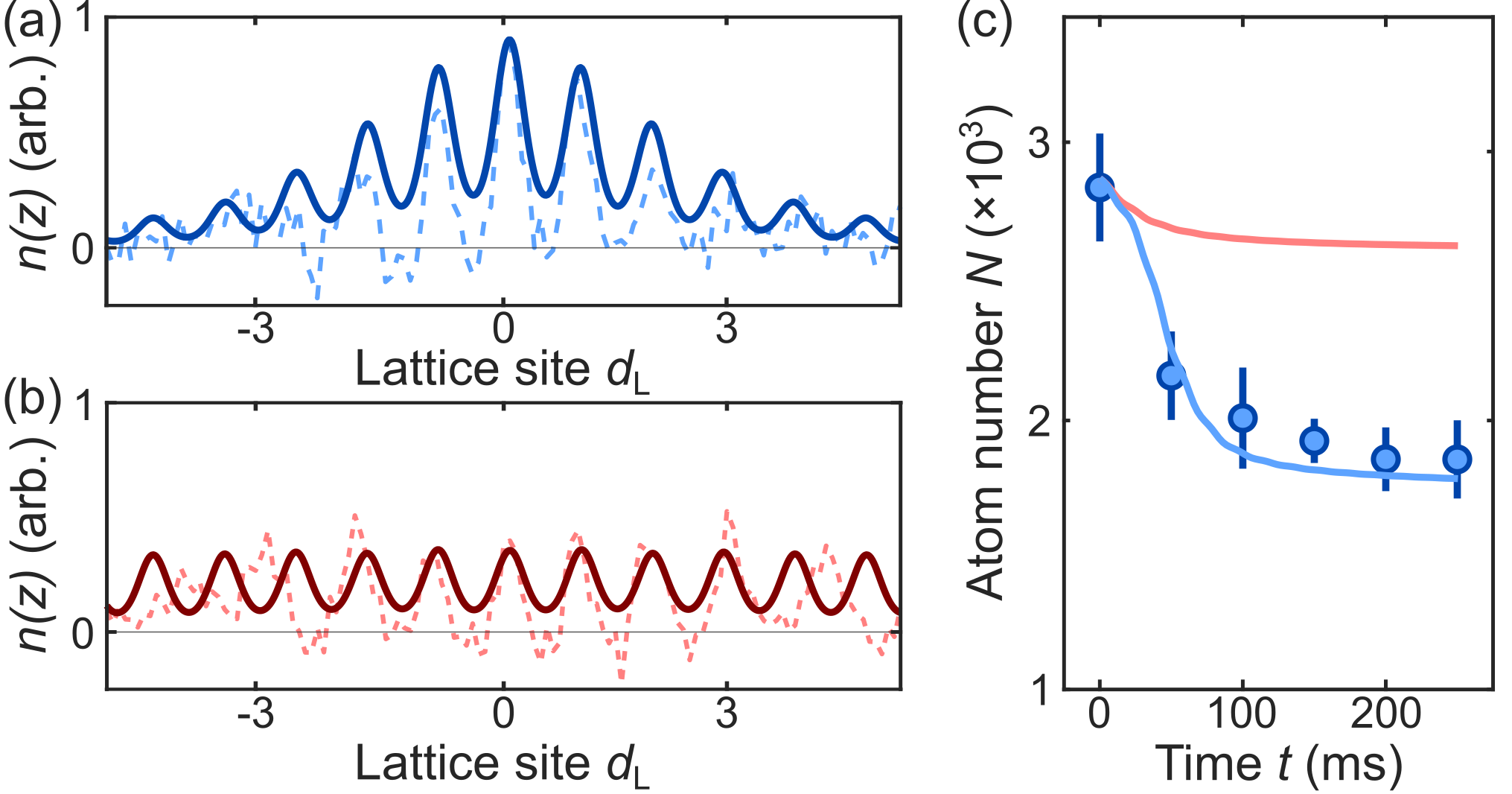}
    \caption{Simulation of the multi-site wave packet. (a) Density profile after $250$\,ms for data in Fig.\,3(b) and 3(c) (blue dotted line) and the numerical simulation (blue solid line) Parameters: $a_s=-5.7\,a_0$, $N_0=2900$, $L_3=5\times  10^{-39}\,$m$^6$s$^{-1}$. (b) Density profile of data in Fig.\,3(a) and 3(c) (red dotted line) and simulation with $a_s=+2\,a_0$ (red solid line). (c) Observed atom number in Fig.\,3(d) (blue circles) and the atom numbers resulting from the simulations in (a) and (b) (blue and red lines respectively). \label{fig:sim_multisite} }    
\end{figure}

{\bf Simulation of the multi-site wave packet.} 
We used the same approach to simulate the time-evolution of the density distribution shown in Fig.\,3. To initialize the simulation, we matched the site occupation numbers of the initial state to our measurements and propagated the state for $250$\,ms using Eq.\,(1). The results show good agreement between the numerical simulation [solid line in Fig.\,\ref{fig:sim_multisite}(a)] and the experimental data (dashed line) for $a_s=-5.7\,a_0$. The occupation data is taken after a magnification ramp that preserves the site-resolved occupation numbers but introduces two effects. It increases $d_\text{L}$ to $20(1)\,\upmu$m and reduces the atomic density between lattice sites. Additionally, the ramp can induce small oscillations of atoms within each site, resulting in minor shifts of the observed peak positions.

The same simulation procedure was applied for the case of a scattering length of $+2\,a_0$  for the measurement in Fig.\,\ref{fig:sim_multisite}(b). Once again, we observe good agreement in the spreading behavior between the simulation [solid red line in Fig.\,\ref{fig:sim_multisite}(b)] and the experiment (dashed red line). The slight reduction in the measured atom number at the outermost sites is again attributed to our magnification scheme, which limits spatial detection due to the finite extent of the laser beams.

Finally, we also found good agreement in the total atom number obtained from the simulation [blue line in Fig.\,\ref{fig:sim_multisite}(c)] with the corresponding experimental measurements [blue circles in Fig.\,\ref{fig:sim_multisite}(c)]. We found that the atom loss is substantially reduced when the atomic density decreases for a spreading wave packet [red line in Fig.\,\ref{fig:sim_multisite}(c)].

\bibliography{References}

\end{document}